# Originality in scientific titles and abstracts can predict citation count

Jack H. Culbert[*,1], Yoed N. Kenett[2] and Philipp Mayr[1]

*jack.culbert@gesis.org*

[1] GESIS – Leibniz Institute for the Social Sciences, Unter Sachsenhausen 6-8, Cologne, Germany
[2] Faculty of Data and Decision Sciences, Technion - Israel Institute of Technology, Kiryat Hatechnionm, 3200003, Haifa, Israel

**Abstract**
In this research-in-progress paper, we apply a computational measure correlating with originality from creativity science: Divergent Semantic Integration (DSI), to a selection of 99,557 scientific abstracts and titles selected from the Web of Science. We observe statistically significant differences in DSI between subject and field of research, and a slight rise in DSI over time. We model the base 10 logarithm of the citation count after 5 years with DSI and find a statistically significant positive correlation in all fields of research with an adjusted $R^2$ of 0.13.

**Introduction**

One aspect of abstracts that likely varies across scientific domains and changes over time is the abstract originality. While some scientific domains have strict norms on abstract formats and content, the increased challenge of a scientific paper getting attention, due to rapid increase in volume of papers with decreased attention span due to information overload (Hołyst, et al., 2024), likely impacts the originality of abstracts. However, the impact of such pressures on abstract writing could have both a facilitative or inhibitory impact on their originality: Abstracts may become more original over time, to compete for a reader's attention more strongly, or they may become less original, to standardize within scientific disciplines and minimize information overload. A possible way to examine these competing hypotheses is by harnessing computational tools that have been recently developed to quantitatively assess the originality of short narratives, particularly an approach called Divergent Semantic Integration.

Cognitive research developed alongside linguistics and natural language processing (NLP) research, as one of the original goals of NLP was to develop a "general theory of human language understanding" which is "linguistically meaningful and cognitively plausible" (Lenci & Padó, 2022). Recent advancements in NLP over the last 10 years have continued to be utilized in modern cognitive research, aided by the rapid development of (large) language models based on deep learning techniques, in particular transformer models.

Divergent Semantic Integration (DSI) (Johnson, et al., 2023) is a computational metric for short textual narratives which was shown to correlate with empirical measures of originality. DSI is computed as the arithmetic mean of cosine distances between embeddings of sentences from a language model, measuring the overall richness of the language used by the writer in their narrative.

The driving concept is that divergent ideas contained within the text are mapped to distant areas within the embedding space of the model, thereby more diverse concepts are more distant to each other on average than similar or uncreative concepts – resulting in a higher DSI score. Extensive empirical creativity research has highlighted how higher creative individuals exhibit a richer memory structure and are able to more broadly search, expand, and create original ideas (Beaty & Kenett, 2023) (Benedek, Beaty, Schacter, & Kenett, 2023).

This study follows previous research into creativity in science, which has mainly focused on a research paper's metadata, for example: the age of keywords (Azoulay, Zivin, & Manso, 2011), novel or unusual combinations of keywords (Boudreau, Guinan, Lakhani, & Riedl, 2016), referenced articles (Trapido, 2015) or the network centrality between citing and cited papers, (Shibayama & Wang, 2020), the lattermost notably was also found to correlate with citations.

In this study, we compute the DSI of the combined titles and abstracts of papers contained within Clarivate's Web of Science (WoS) from a diverse number of fields and over time, to explore whether there exist trends in originality that correlate with field of research, primary subject classification, bibliometric measures, publication date, or citation count.

**Methodology**

DSI is computed as the arithmetic mean of the pairwise cosine distance of the embeddings (produced by BERT (Devlin, Chang, Lee, & Kristina, 2019) in layers 6 and 7) of the sentences in a text with each other. The cosine distance is defined as one minus the inner product of the two input vectors. Equivalently this is formulated as, for a text $T$ defined as an ordered list of length $n > 2$ containing sentences $s_i$, and the embedding vector from the BERT model at layer $k$ represented as $BERT_k(s_i) = \beta_{i,k}$:

$$DSI([s_1, s_2, \ldots, s_n]) = \sum_{k_1, k_2 \in \{6,7\}} \sum_{1 \leq i < j \leq n} \frac{1 - \frac{\beta_{i,k_1} \cdot \beta_{j,k_2}}{\|\beta_{i,k_1}\| \|\beta_{j,k_2}\|}}{4n}$$

We based our code on the codebase provided alongside (Johnson, et al., 2023) and applied this to the combined title and abstract of articles in a snapshot of the WoS. We augmented the original code through refactoring it into a vectorised function that can be applied in a distributed manner against the databases. We computed the DSI of the titles and abstracts, as detailed in the Data section, and then performed a statistical analysis of the DSI against the other variables as detailed in the Results section.

*Data*

In this study, we obtained the abstracts and bibliometric information from the WoS as of April 2024, provided by the Competence Network for Bibliometrics.[i] From this database we retrieved all subjects with over 10,000 records with classification "Article". Of these we chose subjects which have at least 1000 abstracts with 199-299 spaces, which we assumed correlates to 200-300 words in each abstract. This sampling strategy was chosen to accommodate the long computation time that DSI requires, and to allow for easier analysis of the data.

As mentioned in the discussion of Figure 2 we did not select an equal number of papers per year, which led to an underrepresentation of older papers–for our continuing work we will resample with an even distribution of papers per year and compute the DSI scores for this new dataset.

After this filtering we arrived at a dataset with 1238 candidate subjects, corresponding to approximately 1,238,000 articles, which is ~1.65% of the WoS. After evaluating the scalability of the code, we observed that an abstract of the required length took around 18.2 minutes (after improving the performance of the code), which was mainly attributable to the asymptotic quadratic complexity of computing the pairwise cosine distance over all embeddings generated in the DSI computation.

We took the largest 100 subjects by paper count since 1980 in the WoS and chose a random sample of 1000 articles with 200-300 words in their abstract, these were not balanced to be representative of the number of papers published by year. We appended the abstract to the title (with a full stop in-between) and used this to compute the DSI for each article, ending in a dataset of 100,000 abstracts analysed.

Furthermore, we removed all 443 articles from 2024 from the analysis, as the April edition of the WoS had collected an unrepresentatively small sample for 2024 in the months before the snapshot. This left us with a final dataset of 99,557 records to analyse.

Alongside the DSI scores the following bibliometric information was extracted from the Competence Network for Bibliometrics' version of the WoS: "Primary Subject", "Publication Year", "Citations after 3 Years", "Citations after 5 Years" and "Total Citations". We identified the field of research (field) for each primary subject through correlating with CWTS' NOWT classification[ii] and Clarivate's Research Areas,[iii] which is visible in Figure 3. Notably in the NOWT classification, the subject Multidisciplinary Sciences was classified into its own field, and we follow this convention, although this leads to a comparatively higher variance for this field due to its smaller size.

**Results**

The distribution of DSI by fields of research is plotted in Figure 1 (left). We observe a broadly symmetric distribution around the mean for each field, with long tails. We also note a small difference in mean DSI between fields and a similar range to each field. Performing an ANOVA F-test on these categories resulted in statistics $F(5, 99551) = 5936$, $p < 0.01$, $\eta^2 = 0.298$, confirming that the categories have statistically significant differences in means at a 99% confidence level.

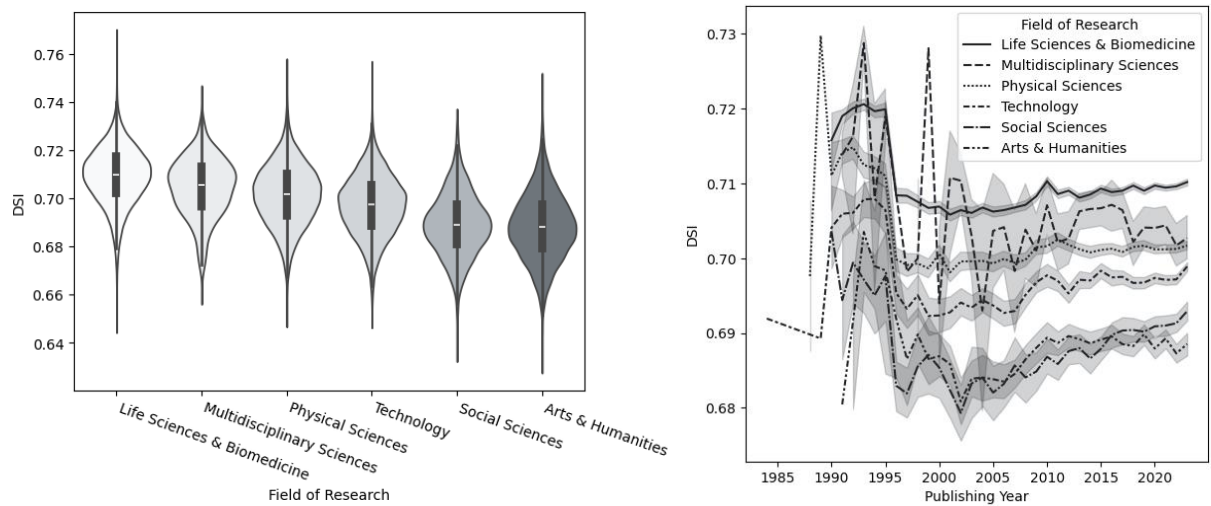

**Figure 1. (Left) Violin plots of the DSI for each field, ordered by mean DSI.**
**(Right) Line plots of DSI by publication year with 95% confidence interval, by field.**

Observing the progression of DSI per field over time in Figure 1 (right),Figure 2 we see a higher average DSI in the 1990s, which falls and remains stable if not trending slightly positive since 1997 for each field excluding Multidisciplinary Sciences.

Following this observation, we investigated the higher mean and variance of DSI prior to 1997. We found an imbalance of records in our dataset by year–following the well reported global rise in number of papers published by year–which led to an underrepresentation of records the earlier that they were published, due to our random sampling strategy. As mentioned previously, the data will be resampled for following work to correct for this bias.

We modelled citation count using a multilinear model of DSI and field as a categorical variable. We mitigated the bias due to accrual of citations by older papers by correlating the number of citations after 5 years, so for this model we considered only papers published before the end of 2018, to allow for a fair accrual of 5 years of citations before the 2024 sample date. This restriction left us with a dataset of 64,816 records.

As some subjects had a large range in citation count after 5 years, and to better model the large differences in average citation count after 5 years by subject, we took the base 10 logarithm of the citation count after 5 years, (after adding 1 to all citation counts to prevent logarithm errors for papers with no citations).

In Figure 2 we observe a positive correlation between the DSI and base 10 logarithm of the citation count after 5 years for all fields. We performed a statistical analysis of the model: $log_{10}(cit_{5\ years} + 1) \sim DSI + C(Field)$, which was found to be statistically significant by two-tailed hypothesis test at 99% confidence. The model has a MSE of 0.24, adjusted $R^2$ of 0.130, Jarque-Bera of 12.918, and a skew and kurtosis of 0.022 and 2.947 respectively. This implies the model explains ~13% of the variation in citation counts. The model may be improved by incorporating publishing year, author count or other bibliometric information, however due to the nature of citation behaviour and the limitations of only analysing titles and abstracts we do not expect a significantly stronger model.

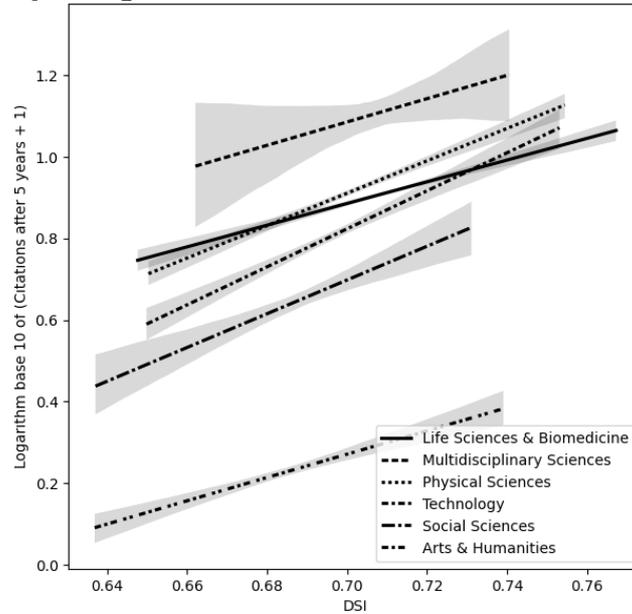

**Figure 2. Least Squares Regression for base 10 of the number of citations after 5 years (plus one) as predicted by DSI and field, plotted with 95% confidence interval.**

In Figure 3 we break down fields to primary subjects and plot the DSI as a bar chart. We observe broadly similar distributions in DSI across subjects: a unimodal bell-curve with thin, long tails and large overlap of the distribution of DSI between subjects and fields.

In our dataset the five subjects with highest mean DSI in descending order are: Cardiac & Cardiovascular Systems (μ = 0.717, σ = 0.00932), Ophthalmology (μ = 0.715, σ = 0.0110), Gastroenterology & Hepatology (μ = 0.715, σ = 0.00986), Urology & Nephrology (μ = 0.714, σ = 0.00942) and Obstetrics & Gynecology (μ = 0.714, σ = 0.0111).

The five subjects with lowest mean DSI in descending order are: Philosophy (μ = 0.687, σ = 0.0137), Education & Educational Research (μ = 0.686, σ = 0.0122), Art (μ = 0.686, σ = 0.0132), Political Science (μ = 0.686, σ = 0.0116) and History (μ = 0.683, σ = 0.0136).

**Discussion and Conclusion**

This large-scale (n = 99,557) ongoing study of the DSI of abstracts and titles in the Web of Science was intended to explore whether this metric, demonstrated in (Johnson, et al., 2023) to be correlated with originality of narratives, also correlates with bibliometric variables.

Our most significant finding so far in this study is our modelling of the logarithm of citation counts after 5 years by DSI and field, which resulted in statistically significant positive correlations which indicate DSI may be a useful computational indicator for future citations (Figure 2).

We observed a statistically significant difference in DSI by field of research, as well as a slight positive trend over time. As there is a large overlapping spread of DSI between fields, this implies that categorising subjects by field may not be the best discriminator for DSI.

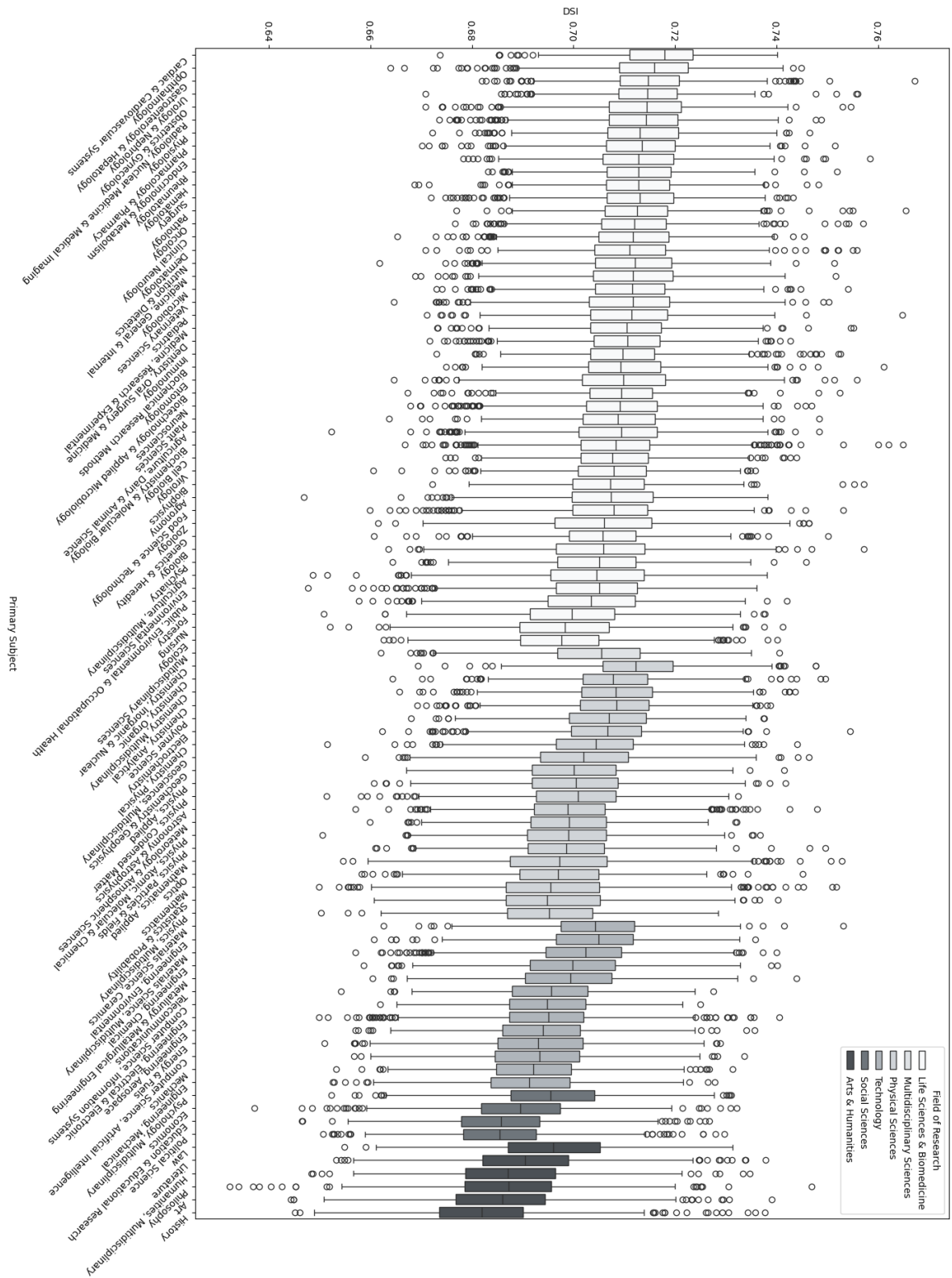

**Figure 3. Boxplot of DSI scores per subject and field, ordered by mean DSI including outliers and plotted with mean excluding outliers.**

We note that subjectively, technologically applied fields appear to have higher DSI than less technologically applied fields. This may be due to the tokenisation and embedding of novel terms creates vectors that do not align with the rest of their field (potentially due to the lack of exposure for the model in training), therefore a next step would be to experiment with a model trained on scientific text such as SciBert (Beltagy, Lo, & Cohan, 2019) for this analysis.

A fundamental limitation of our study is the lack of human-ranked creativity scores for scientific papers, and our assumption that DSI generalises past to scientific ones as a metric of originality. As mentioned previously, our dataset was not balanced in terms of publishing year, which diminishes the strength of our findings in the positive trends of DSI mapped over time.

Furthermore, while DSI was found to generalise across varying language and cultural backgrounds in study 6 of (Johnson, et al., 2023), we have not controlled for English proficiency in this study. Similarly, in study 5 DSI was found to stabilise after 30-50 words up to 200 and was not evaluated at the length we are considering at approximately 200-300 words.

We look to extend this study through analysis of a new collection of data, further analyses of the correlation of DSI with other bibliometric indicators available and computed in the Competence Network for Bibliometrics' version of the Web of Science database to refine our modelling of DSI, as well as experimenting with the embedding model for DSI.

Our results indicate a promising content-based computational method for analysis of scientific papers and potentially a novel link between the creativity sciences and Scientometrics. Computational measures such of these may be of use to the bibliometric community in the analysis of creativity and originality in papers, and perhaps for the wider academic community if this or other originality metrics are incorporated into a search engine as an additional index to re-rank retrieved items.

---

[i] https://bibliometrie.info

[ii] https://www.cwts.nl/pdf/nowt_classification_sc.pdf

[iii] https://images.webofknowledge.com/images/help/WOS/hp_research_areas_easca.html